\documentstyle[12pt]{article}

\newcommand{\be}{\begin{equation}}
\newcommand{\ee}{\end{equation}}

\newcommand{\bea}{\begin{eqnarray}}
\newcommand{\eea}{\end{eqnarray}}

\newcommand{\p}{\partial}

\newcommand{\nn}{\nonumber \\}
\newcommand{\f}{\frac}

\begin{document}
\thispagestyle{empty}

\begin{flushright}
{\bf arXiv: 1003.0745}
\end{flushright}
\begin{center} \noindent \Large \bf
Weak Gravity Conjecture, Central Charges and $\eta/s$ 
\end{center}

\bigskip\bigskip\bigskip
\vskip 0.5cm
\begin{center}
{ \normalsize \bf  Shesansu Sekhar Pal}

\vskip 0.5cm

\vskip 0.5 cm
Center for Quantum Spacetime, \\
Sogang University, 121-742 Seoul, South Korea\\
\vskip 0.5 cm
\sf shesansu${\frame{\shortstack{AT}}}$gmail.com
\end{center}
\centerline{\bf \small Abstract}
We correlate the weak gravity conjecture (WGC), the  KSS conjecture with chemical potential at extremality and the central charges by going through a particular  example in five dimensional AdS spacetime with two unknown coefficients $c_1,~c_2$, assuming  WGC exists in AdS spacetime. 
The result  that follows from this example suggests that WGC  makes  the KSS conjecture to hold  in the extremal limit but only when one of the coefficient vanishes ($c_1=0,~c_2\neq 0$ or  $c_2=0,~c_1\neq 0$)  and when both the coefficients are non zero  it  can respect and/or violate the KSS conjecture depending on the choice to $c_1$  at extremality, even though $\eta/s$ do not depend on $c_1$ at extremality.   
Moreover, WGC is not fully compatible with the calculation of central charges even though the  bounds on  coefficient $c_1$ that follows from demanding WGC stays within the bounds that central charges predict. As usual, the KSS conjecture is violated, of course, in the non-extremal limit.
\newpage

\section{Introduction}

The weak gravity conjecture (WGC) in flat spacetime puts a very important restriction on the ratio of mass or energy density of particles or black holes to the charge it carries, in appropriate units it should  be less than or equal to unity \cite{ahmnv}. On assuming that there exists a similar kind of WGC in the AdS spacetime puts some interesting restrictions on the coefficients that appear in the
low energy effective
 gravitational action in the finite 't Hooft coupling limit. The application of the holographic correspondence \cite{ads_cft} to   systems which is described  by that kind of  action yields interesting connections between WGC and holography. In particular, we shall explore the connection between    KSS conjecture \cite{kss} at extremality and the WGC. 

The central charges of the dual field theory are connected to these coefficients that appear in the action of the effective bulk theory and hence appear in the computation of $\eta/s$ of the plasma.

Hence, it is natural to think  all these three: WGC, KSS conjecture and the central charges are all correlated. In this paper we are going to demonstrate this by considering  a specific example in the extremal limit.  We expect the result, especially the connection between the WGC and the KSS conjecture is generic in the extremal limit.

\subsection{Conjectures: WGC and KSS}

The WGC \cite{ahmnv}, came out of the fact that  gravity is the weakest force and says that a stable charged particle minimizes the ratio of energy density to charge density and is less than unity in some units\footnote{Here we are using the terminology, energy density for mass and charge density for charge.}. Importantly, as emphasized in \cite{ahmnv} the conjecture is not for a given a charge sector for which the masses are less then the charge but there could exists  
some states of this type. 

It has also been suggested in \cite{ahmnv} that WGC came out of the requirements of having finite number of stable particles which are not protected by any symmetry principle and fits in nicely with the absence of having any global symmetries in a consistent quantum gravitational theory. 

The outcome of the WGC is that for extremal black hole with the appropriate ratio of energy density to charge density becomes unity and the ratio can go below unity for small charge corrections.

In \cite{ahmnv}, the authors suggested two different forms of the conjecture, one for the state with lightest charged particles and the other for the state with smallest mass to charge ratio. The combination of these two forms suggests  the existence of lightest charged particle with mass to charge ratio should be less than that of the  mass to charge ratio of the extremal black hole.  
In this paper we shall adopt this as our  guiding principle and find out the consequence of this. 

Probably, it is correct to say that WGC holds for arbitrary  rank of the gauge group  and 't Hooft coupling, even though we will be restricting ourselves only to abelian gauge groups. Also, it does not depend on the nature of the asymptotic spacetime or the dimension of spacetime. We shall assume this and proceed further and  apply it to situations where we are applying the holographic principle.

It is important to note that  the holographic ways to calculate  central charges, thermodynamics quantities or  the transport coefficients always receive corrections from both the finiteness of the rank of the gauge group and 't Hooft coupling.

The KSS conjecture \cite{kss} suggests that the ratio of coefficient of shear viscosity $\eta$ to entropy density $s$ must have a minimum value of $1/4\pi$ at zero chemical potential for theories that admits gravity dual. Of course, we have already witnessed several examples of the violation of it \cite{ssp}-\cite{others}, at finite 't Hooft coupling. But have not seen an example where it can be violated in the extremal limit at finite 't Hooft coupling.

It is easy to convince oneself that  $\eta/s$  attains the lower bound in the extremal limit for Einstein-Maxwell type of  theories described by actions having only two derivatives \cite{ejl}. This is due to the fact that  any  charged $AdS_d$ spacetime in the  extremal limit becomes that of $AdS_2\times R^{d-2}$ spacetime and we already know the result to $\eta/s$ that comes from pure AdS spacetime. But whenever, there is an interaction between the U(1) gauge fields and  metric, which occurs for theories with more than two derivatives,  then there is no reason to believe that the  result to $\eta/s$ should obey the KSS conjecture even at extremality. However, due to WGC,  one can show that   $\eta/s$  obeys the KSS conjecture  in the extremal limit in some cases  and violates in some other cases. 
 
The philosophy and the plan that we shall adopt in this paper consists of three steps.

Step 1: We shall look for the ratio of energy density to charge density for the given gravitational system in five spacetime dimension with unknown coefficients $c_i$'s and then find the restriction on these coefficients that follows from demanding this ratio to be smaller than unity in appropriate units\footnote{This step was inspired by an analogous calculation done in asymptotically flat spacetime \cite{kmp}.}.

Step 2: We shall use the result of the calculation of the central charges \cite{hm} to fix some of the coefficients $c_i$'s and use  step one to put restriction on rest of the coefficients $c_j$'s.

Step 3: We shall calculate the ratio  $\eta/s$ in the extremal limit and examine  what happens to the KSS conjecture if we take the restriction that follows from WGC in the first step as well as those that follows from central charges in step two.

The result of this study can be summarized as follows. We consider a low energy effective gravitational theory with two   unknown coefficients $c_1$ and $c_2$ and we  fix one of the coefficient $c_1$ with the central charges of the corresponding dual field theory following the holographic anomaly calculation \cite{no}. Then we use the result of \cite{hm} to get the bounds on this coefficient, and the result suggests it can take both positive and negative values. Now, demanding WGC, we get the restriction on the other coefficient $c_2$ which too can take both positive and negative values. The $\eta/s$ in the extremal limit depends only on one coefficient that is on $c_2$, 
which means the KSS conjecture is violated even in the extremal limit only in the case where both $c_1$ and $c_2$ are non zero. But when $c_1=0$, and  $c_2$ is not  WGC make KSS conjecture to hold at extremality. When $c_2=0$ and $c_1$ is not WGC rather contradicts with the result of the central charge calculation of \cite{hm}, but the KSS conjecture at extremality survives.

The paper is organized by  assuming a specific  form of  the low energy effective action. Then we study the thermodynamics and the transport properties of the black hole solution using the known recipes. Then we go through the three steps as mentioned above to find out the correlation among WGC, KSS conjecture and the central charges. In appendix, we calculate $\eta/s$ for another example and show that it is independent of the coefficient that appear in the higher derivative term to metric but depends on the coefficient that appear in the interaction term between the gauge field and the metric degrees of freedom.

\section{A gravitational system}

Let us consider an effective action of the following type
\be\label{action}
S=\f{1}{2\kappa^2}\int \sqrt{-g}[R+12-\f{1}{4}F^{MN}F_{MN}+c_1R_{MNKL}R^{MNKL}+c_2 R^{MNKL} F_{MN}F_{KL}],
\ee
this action is a special case to the action considered in \cite{mps}, but for our purpose this is good enough. 

It admits the following form of the black hole solution
\bea\label{ansatz}
ds^2&=&-r^2a(r)^2dt^2+r^2 (dx^2+dy^2+dz^2)+\f{dr^2}{r^2b(r)^2},\nn
A&=&h(r)dt,
\eea
with 
\bea
a(r)^2&=&\bigg[1 - \bigg(\f{r_0}{r}\bigg)^2\bigg] \bigg[1 - \f{q^2}{r^4 r^2_0} + \bigg(\f{r_0}{r}\bigg)^2\bigg] + 
 c_2 \bigg[-\f{4 q^4}{r^{12}} - \f{40 q^2}{r^6} - \f{4 q^4}{r^4 r^8_0} + \f{32 q^2}{
    r^4 r^2_0} +\nn&& \f{8 q^2 r^4_0}{r^{10}}\bigg(1 + \f{q^2}{r^6_0}\bigg) \bigg] + 
 c_1 \bigg[\f{17 q^4}{6 r^{12}} - \f{158 q^2}{3 r^6} - \f{23 q^4}{2 r^4 r^8_0} + \f{
    42 q^2}{r^4 r^2_0} - \f{2 r^4_0}{r^4} + \f{20 q^2r^4_0}{
    3 r^{10}} \bigg(1 + \f{q^2}{r^6_0}\bigg)  + \nn && \f{2r^8_0}{r^8} \bigg(1 + \f{q^2}{r^6_0}\bigg)^2 \bigg],\nn
b(r)^2&=&\bigg[1 - \bigg(\f{r_0}{r}\bigg)^2\bigg] \bigg[1 - \f{q^2}{r^4 r^2_0} + \bigg(\f{r_0}{r}\bigg)^2\bigg]  + 
 c_1 \bigg[\f{17 q^4}{6 r^{12}} - \f{158 q^2}{3 r^6} - \f{23 q^4}{2 r^4 r^8_0} + \f{
    42 q^2}{r^4 r^2_0} - \f{2 r^4_0}{r^4} +\nn
&& \f{20 q^2r^4_0}{
    3 r^{10}} \bigg(1 + \f{q^2}{r^6_0}\bigg)  + \f{2r^8_0}{r^8} \bigg(1 + \f{q^2}{r^6_0}\bigg)^2  + 
    \bigg(\f{2}{3}  - 
     \f{104q^2}{3r^6}\bigg) \bigg(1 - \f{r^2_0}{r^2}\bigg) \bigg(1 - \f{q^2}{r^4 r^2_0} + \f{r^2_0}{r^2}\bigg)\bigg] +\nn&&
  c_2 \bigg[-\f{4 q^4}{r^{12}} - {40 q^2}{r^6} - \f{4 q^4}{r^4 r^8_0} + \f{32 q^2}{
    r^4 r^2_0} + \f{8 q^2r^4_0}{r^{10}} \bigg(1 + \f{q^2}{r^6_0}\bigg) 
- \f{
    32 q^2}{r^6} \bigg(1 - \f{r^2_0}{r^2}\bigg) \bigg(1 - \f{q^2}{r^4 r^2_0} + \f{r^2_0}{r^2}\bigg)\bigg],\nn
h(r)&=&\sqrt{3} q \bigg(-\f{1}{r^2} + \f{1}{r^2_0}\bigg) + 
 \f{c_1}{\sqrt{3}} \bigg[-\f{13 q^3}{ r^8} +\f{ q}{ r^2} + \f{13 q^3}{
     r^8_0} - \f{q}{ r^2_0}\bigg] + \nn&&
 c_2 \sqrt{3}\bigg[\f{16  q^3}{r^8} + \f{8  q}{r^2} - \f{8  q^3}{
    r^8_0} - \f{8  q r^4_0}{r^6} \bigg(1 + \f{q^2}{r^6_0}\bigg) \bigg].
\eea

The black hole has a horizon at $r=r_0$, which is the outer horizon, of course there exists an inner horizon, as well. This solution is characterized by electric charge density, $2\sqrt{3}q$ and energy density or mass density $\rho_E$, in units where we have set the AdS radius to unity.  

The solutions readily obeys the restriction that the gauge potential should vanish at the horizon so as to have the vanishing norm at the horizon. The boundary speed is also set to unity as well as the charge density, which is set to $2\sqrt{3}q$.  

\subsection{Thermodynamics}

This is already calculated in \cite{mps} using background subtraction method, but for completeness, we shall record some of the formulae. The free energy density, $w=-\f{S_{BH}-S_{AdS}}{\beta V_3}$, where, $\beta=1/T$, is the inverse temperature associated to the charged black hole, $V_3$ is the volume of the spatial coordinates, $S_{BH}$, is the action of the charged black hole and $S_{AdS}$, is the action of the pure AdS spcetime with a constant gauge potential evaluated with higher derivative correction and when evaluated, it  gives
\be
w=-\f{r^4_0}{2\kappa^2}\bigg[1+\f{19}{3}c_1+\f{q^2}{r^6_0}\bigg(1-\f{113}{3}c_1-32c_2\bigg)+\f{q^4}{r^{12}_0}\bigg(\f{23}{2}c_1+4 c_2\bigg)\bigg]
\ee

The temperature, $T$ and the chemical potential, $\mu=lim_{r\rightarrow \infty} \f{A_0}{\pi}$ are
\bea
T&=&\f{r_0}{\pi}\bigg[1-\f{5}{3}c_1-\f{q^2}{2r^6_0}\bigg(1+\f{31}{3}c_1+16 c_2\bigg)-\f{q^4}{r^{12}_0}\bigg(9 c_1-4 c_2\bigg)\bigg],\nn 
\mu&=&\f{q\sqrt{3}}{\pi r^2_0}\bigg[1-\f{c_1}{3}+\f{q^2}{r^6_0}\bigg(\f{13}{3}c_1-8 c_2\bigg)\bigg]
\eea

Introducing a dimensionless parameter $\bar\mu=\f{\mu}{T}$,  the charge density $n_q=-\bigg(\f{\p w}{\p \mu}\bigg)_T$, which is proportional  to $q$ and the horizon radius $r_0$ are related to $T$ and ${\bar \mu}$ as
\bea
q&=&\f{\pi^3 T^3{\bar \mu}}{\sqrt{3}}[1+\f{{\bar \mu}^2}{3}-\f{{\bar \mu}^4}{36}+8 c_2 {\bar \mu}^2+c_1(\f{11}{3}+\f{26}{9} {\bar \mu}^2+\f{{5\bar \mu}^4}{12})+{\cal O}\bigg({\bar \mu}^5,c^2_1,c^2_2\bigg)], \nn 
r_0&=&\pi T\bigg[1+\f{{\bar \mu}^2}{6}-\f{{\bar \mu}^4}{36}+\f{{\bar \mu}^6}{108}+\f{c_1}{3}\bigg(5+\f{11}{2}{\bar \mu}^2-\f{5}{36}{\bar \mu}^4-\f{23}{108}{\bar \mu}^6\bigg)+\nn&&\f{c_2{\bar \mu}^2}{3}\bigg(8-\f{4}{3}{\bar \mu}^2+\f{4}{27}{\bar \mu}^4\bigg)+{\cal O}\bigg({\bar \mu}^7,c^2_1,c^2_2\bigg)\bigg]
\eea

After re-expressing the free energy density in terms of chemical potential and temperature
\be
w=-\f{\pi^4 T^4}{2\kappa^2}\bigg[1+13c_1+\bigg(1+\f{11}{3}\bigg){\bar\mu}^2+\f{1}{6}\bigg(1+\f{26}{3}c_1+24 c_2\bigg){\bar\mu}^4-\f{1}{108}\bigg(1-15c_1\bigg){\bar\mu}^6 +{\cal O}\bigg({\bar \mu}^7,c^2_1,c^2_2\bigg)\bigg],
\ee

From which it just follows trivially using the identity for entropy density 
\be
s=-\bigg(\f{\p w}{\p T}\bigg)_{\mu}=\f{4\pi^4 T^4}{2 \kappa^2}\bigg[1+\f{{\bar\mu}^2}{2}+\f{{\bar\mu}^6}{216}+c_1\bigg(13+\f{11}{6}{\bar\mu}^2-\f{15}{216}{\bar\mu}^6\bigg)+\cdots\bigg]
\ee

The energy density
\bea
\rho_E&=&w+T(s +{\bar\mu} n_q)=-3w,\nn
&=&\f{3r^4_0}{2\kappa^2}\bigg[1+\f{19}{3}c_1+\f{q^2}{r^6_0}\bigg(1-\f{113}{3}c_1-32c_2\bigg)+\f{q^4}{r^{12}_0}\bigg(\f{23}{2}c_1+4 c_2\bigg)\bigg]
\eea

The extremal limit corresponds to
\be
q^2=2r^6_0[1-48 c_1],
\ee
which is independent of $c_2$. 
On evaluating the energy density in the extremal limit gives
\be\label{ratio_m_q}
\f{2^\f{5}{3}(2\kappa^2)}{9}\f{\rho_E}{q^{\f{4}{3}}}=1-\f{1}{3}\bigg(23 c_1+48 c_2\bigg)
\ee

\subsection{shear viscosity and $\eta/s$ }

The coefficient of shear viscosity can be very easily  computed using the prescription given in \cite{liu} and the result is same as in \cite{mps}
\be
\eta=\f{r^3_0}{2\kappa^2}\bigg(1-24 c_1 \f{q^2}{r^6_0}\bigg)
\ee
 and the entropy density in terms of $q/r^3_0$, can be calculated using the Wald's entropy formula \cite{wald}
\be
s=\f{4\pi r^3_0}{2\kappa^2}\bigg[1+8 c_1-\f{q^2}{r^6_0}\bigg(28 c_1+24 c_2\bigg)\bigg]
\ee

The ratio, $\eta/s$, simply reads to leading order in $c_i$'s as
\be
\eta/s=\f{1}{4\pi}\bigg[1-8 c_1+\f{q^2}{r^6_0}\bigg(4 c_1+24 c_2\bigg)\bigg],
\ee 
which in the extremal limit gives 
\be\label{etas_ext}
\bigg(\f{\eta}{s}\bigg)_{{\rm extremal}}=\f{1}{4\pi}\bigg[1+48 c_2\bigg],
\ee
which is independent of $c_1$, \cite{mps}. It looks like in the extremal limit, it is the coefficient of the interaction term between the gauge field and the metric, which is $c_2$,  that appears in the computation of $\eta/s$. This particular feature of $\eta/s$ also appear in another example that is studied in the appendix.

\section{WGC, KSS conjecture and central charges}

Let us apply the WGC to eq(\ref{ratio_m_q}), i.e. considering the left hand side as the ratio of mass density to charge density means the right hand side must obey 
\be
1-\f{1}{3}\bigg(23 c_1+48 c_2\bigg) \le 1,
\ee 

which  gives the constraint
\be\label{wgc}
0 \le \bigg(23 c_1+48 c_2\bigg) \le 3.
\ee
We shall analyze this equation by considering different cases. 
First, let us consider a simpler case, where $c_1=0$, in this case we can rewrite the constraint as
\be
0 \le 16 c_2 \le 1,
\ee

which simply means the coefficient $c_2$ is positive and the ratio, eq(\ref{etas_ext}) obeys
\be
\f{\eta}{s} \geq \f{1}{4\pi},
\ee
 which  is nothing but respecting the KSS conjecture as the coefficient $c_2$ is positive. So, we just saw the imposition of 
WGC means respecting the KSS conjecture in the extremal limit but only when $c_1=0$. 

For, $c_1\neq 0$, a priori it is not clear why $c_2$ should be positive and hence respect the KSS bound?

There is one more ingredient that we have not yet taken into consideration and that is the central charges. We know from AdS/CFT correspondence that the central charges are related to the anomaly in the one point function of the trace of the energy momentum tensor \cite{hs}.

For the gravity action described by
\be
S=\f{1}{16\pi G}\int d^5x[R+12+\alpha R^2+\beta R^{MN}R_{MN}+\gamma R_{MNKL}R^{MNKL}],
\ee
have the central charges \cite{no}
\bea
\f{c}{16\pi^2}&=& \f{1}{8\pi G}\bigg[\f{1}{16}+\bigg(-\f{5\alpha}{2}-\f{\beta}{2}+\f{\gamma}{4}\bigg)\bigg],\nn
\f{a}{16\pi^2}&=& \f{1}{8\pi G}\bigg[\f{1}{16}+\bigg(-\f{5\alpha}{2}-\f{\beta}{2}-\f{\gamma}{4}\bigg)\bigg],
\eea 
where  the size of the AdS radius has been set to unity. With the choice for which the coefficients are set to $\alpha=0=\beta$ and $\gamma=c_1$, relates to central charges as 
\be
c_1=\f{1}{8}\bigg(1-\f{a}{c}\bigg)
\ee

In an interesting study \cite{hm} have suggested bounds on the ratio of the central charges depending on the amount of supersymmetry preserved
\bea\label{a_c_result}
&&{\rm For}~~~{\cal N}=0,~~~\f{1}{3}~\leq ~\f{a}{c}~\leq~\f{31}{18},\nn
&&{\rm For}~~~{\cal N}=1,~~~\f{1}{2}~\leq ~\f{a}{c}~\leq~\f{3}{2},\nn
&&{\rm For}~~~{\cal N}=2,~~~\f{1}{2}~\leq ~\f{a}{c}~\leq~\f{5}{4},
\eea

Re-writing it in terms of $c_1$, we get the bounds as
\bea\label{bound_c_1}
&&{\rm For}~~~{\cal N}=0,~~~-\f{13}{144}~\leq ~c_1~\leq~\f{1}{12},~~~{\rm imply}~~~\bigg( -\f{299}{432}~\leq ~\f{23}{3}c_1~\leq~\f{23}{36}\bigg),\nn
&&{\rm For}~~~{\cal N}=1,~~~-\f{1}{16}~\leq ~c_1~\leq~\f{1}{16},~~~{\rm imply} ~~~\bigg(-\f{23}{48}~\leq ~\f{23}{3}c_1~\leq~\f{23}{48}\bigg),\nn
&&{\rm For}~~~{\cal N}=2,~~~-\f{1}{32}~\leq ~c_1~\leq~\f{1}{16},~~~{\rm imply}~~~ \bigg(-\f{23}{96}~\leq ~\f{23}{3}c_1~\leq~\f{23}{48}\bigg),
\eea
  
It is interesting to see that the magnitude of $c_1$, which is small and less than unity  all the time and is consistent with our supergravity analysis, but can take  both positive and negative values.

Let us use these constraint that we obtained from the calculation of central charges on the coefficient $c_2$, i.e. eq(\ref{bound_c_1}) in eq(\ref{wgc})
\be\label{wgc_c2}
-23 c_1\leq 48 c_2\leq 3-23 c_1
\ee
and it just follows that $c_2$ is  not always positive, and is  independent of the amount of supersymmetry preserved, which implies from eq(\ref{etas_ext}) that the KSS conjecture is not obeyed all the time even  in the extremal limit. It would be very interesting to cross check this constraint on $c_2$ from direct calculation of anomaly
or from some other ways. See appendix for the full solution of eq(\ref{bound_c_1}) and eq(\ref{wgc_c2}).

It looks like the low energy effective action eq(\ref{action}),  may not be part of the swampland \cite{cv}, as we have taken the WGC condition into account. However, it would be very interesting to consider an effective action which is part of the swampland and calculate  $\eta/s$ in the extremal limit and see what it has got to say about the  KSS conjecture in the extremal limit.

Let us look at the case for which $c_2=0$, then the resulting constraint that follows from eq(\ref{wgc}) is not fully compatible with what follows from the calculation of the central charges i.e. from eq(\ref{bound_c_1}). However, it says that the restriction from WGC stays within the bounds resulting from the calculation of central charges.
 \footnote{If we assume WGC is exact, then it says that the left hand side of eq(\ref{bound_c_1}) need to be corrected and should be set to zero, which would be very interesting to explore. However,  if we assume eq(\ref{bound_c_1}) is not going to be corrected then it says that mass or energy density can be bigger than the charge density. }
It is important to note that in the extremal limit $\eta/s$ do not depend on the coefficient $c_1$, so the  KSS conjecture at extremality is safe. Moreover, in this limit, ($c_2=0$), WGC has got nothing to say on $\eta/s$. 

\section{Conclusion and discussion}

In this paper we have studied the  KSS conjecture with chemical potential in the extremal limit with higher derivative correction and showed by going through an example that it is the WGC \cite{ahmnv} along with the restriction of central charges in four dimensional field theory \cite{hm}, that follows from holography,  makes the ratio of $\eta/s$ to stay above $1/4\pi$  in the extremal limit but only in certain cases. The bounds on the coefficient $c_1$ that  results after imposing WGC do not completely agree with the bounds that results from central charge calculations. In particular, when $c_2=0$, WGC imply $c_1$ should always be positive but central charge calculation \cite{hm} suggests $c_1$ can take negative values.

We have also shown  (also in \cite{mps}) that in the extremal limit the ratio $\eta/s$ depends only on the coefficients $c_2$ that appear in the interaction of the U(1) gauge fields and the metric. Moreover, the condition at extremality which is  the ratio of charge density to the size of the horizon do not depends on the coefficient $c_2$.

It is certainly very interesting to study more examples in the Einstein-Maxwell sector with higher derivative terms as well as going  beyond  Einstein-Maxwell type of examples and examine what happens to  the  KSS conjecture at extremality. It would also be nice to have a bound on $c_2$ without invoking WGC and if it stays positive then the KSS conjecture holds at exremality.

In another context it is not a priori clear which kind of low energy effective action that one needed to consider and what are the criteria(s) to fix the form of such  actions ?
Is it just the symmetry principle that is enough to fix the form ? Or we need to take into account all: WGC, holography  and symmetry principle as the guiding principle to fix the form, which would certainly be very interesting to explore.  
\\

{\bf Note added}: After submitting the paper to arXiv, we are informed  of the paper \cite{cls}, which also discusses the relation between WGC and $\eta/s$.

\section{Acknowledgment}
It is a pleasure to thank Ofer Aharony for going through the manuscript and giving few suggestions, also would like to thank Bum-Hoon Lee for a discussion  and the members of  CQUeST for  their help.

This work was supported by the Korea Science and Engineering Foundation (KOSEF) grant funded by the Korea
government (MEST) through the Center for Quantum Spacetime (CQUeST) of Sogang University with grant number R11-2005-021.

\section{Appendix}

In this appendix, we shall calculate the $\eta/s$ for a 
 low energy effective  action in five dimension, whose form  is similar in nature to \cite{hm}
\be
S=\f{1}{2\kappa^2}\int \sqrt{-g}[R+12-\f{1}{4}F^{MN}F_{MN}+c_1~{\cal C}_{MNKL}{\cal C}^{MNKL}+c_2 ~{\cal C}^{MNKL} F_{MN}F_{KL}],
\ee
where ${\cal C}_{MNKL}$ is the Weyl tensor. In order to calculate $\eta/s$ to leading order in $c_i$'s we do not need to know the full solution, as the metric components along the spatial directions do not receive any corrections to leading order in $c_i$, hence only the zeroth order solution in $(c_1)^0$ and $(c_2)^0$ is good enough.

Let us take the solution to have the form 
\be
ds^2=-r^2a^2(r) dt^2+c^2(r)(dx^2+dy^2+dz^2)+\f{1}{r^2b^2(r)} dr^2,
\ee

with the lowest order solution 
\bea
a^2(r)=b^2(r)&=& \bigg[1 - \bigg(\f{r_0}{r}\bigg)^2\bigg] \bigg[1 - \f{q^2}{r^4 r^2_0} + \bigg(\f{r_0}{r}\bigg)^2\bigg],~~~c^2(r)=r^2 ,\nn
h(r)&=&\sqrt{3} q \bigg(-\f{1}{r^2} + \f{1}{r^2_0}\bigg) 
\eea

This solution has a horizon at $r=r_0$, the gauge potential vanishes at the horizon in order to have a vanishing norm there, the speed on the boundary has been set to unity and the charges is set to the value, proportional to $q$.

The shear viscosity can
be evaluated by computing the following quantity, following \cite{liu}
\be\label{definition_eta}
\eta=\lim_{k_{a}\rightarrow 0} \f{\prod(r,k_a)}{i\omega \phi(r,k_a)}
\ee
at the boundary, where $\prod$ is the momentum associated to  the field $\phi$, which  is related to the  metric fluctuation, Let us denote the metric fluctuation
\be
{h^y}_x=\int[dk] \phi_k(r) e^{-i\omega t+i k z},
\ee
where the graviton is moving along  $z$ direction and we are using a short hand notation to write the appropriate measure
factor for  momentum integrals and factors of $2\pi$ in $[dk]$. The equation of motion can be derived from the following effective action
\bea
S&=&\f{1}{2\kappa^2}\int dr [dk]\bigg(A(r)\phi''_k\phi_{-k}+B(r)\phi'_k\phi'_{-k}+C(r)\phi'_k\phi_{-k}+D(r)\phi_k\phi_{-k}+
\nn &&E(r) \phi''_k\phi''_{-k}+F(r)\phi''_k\phi'_{-k} \bigg)+{\cal K},
\eea
where ${\cal K}$ is the appropriately generalized Gibbons-Hawking boundary term \cite{bls}.

Ignoring the details, the coefficient of shear viscosity, $\eta$
becomes
\be\label{eta}
\eta=lim_{\omega\rightarrow 0}
\f{\prod}{i\omega\phi}=\f{1}{\kappa^2}[\kappa_2(r)+\kappa_4(r)]_{r=horizon}, \ee where \be
\kappa_2=\f{1}{(r^2ab)}\bigg[A-B+\f{F'}{2}\bigg],~~~\kappa_4=\f{d}{dr}\bigg[E\f{d}{dr}\bigg(\f{1}{r^2ab}\bigg)\bigg]
\ee

and on evaluating it 
\bea
\eta&=&\f{c^3}{2\kappa^2}+ \f{c_2}{2\kappa^2} \f{b^2 c^3 h'^2}{a^2}+\f{c_1}{2\kappa^2} \f{c}{a^2} \bigg[-2a^2b^2c^2-2ra^2bc^2b'+2r^2a^2b^2c'^2-\nn &&
2rabc(rca'b'+b(3ca'+ra'c'+rca''))+2r^2a^2bc(b'c'+bc'')\bigg]
\eea

On evaluating the  entropy density using  Wald's entropy formula \cite{wald}
\bea
s&=&\f{2\pi c^3}{\kappa^2}-\f{c_2}{\kappa^2}\f{2\pi b^2 c^3h'^2}{a^2}-\f{c_1}{\kappa^2}\f{2\pi c }{a^2}\bigg[2a^2b^2c^2+2ra^2bc^2b'+2r^2a^2b^2c'^2+\nn &&
rabc(2rca'b'+b(-2ra'c'+2c(3a'+ra'')))-
2ra^2bc(rb'c'+b(2c'+rc''))\bigg]
\eea

The temperature to the zeroth order in $(c_1)^0$ and $(c_2)^0$
\be
T=\f{r_0}{\pi}\bigg[1-\f{q^2}{2r^6_0}\bigg].
\ee
 
This implies  extremality is at
\be
q^2=2~r^6_0~+{\cal }O(c_1,~c_2).
\ee

Using the zeroth order solution, the coefficient of shear viscosity becomes
\be
\eta=\f{ r^3_0}{2\kappa^2}\bigg[1+4 c_1+ \f{q^2}{r^6_0}(-14c_1+12c_2) \bigg]
\ee

and the entropy density becomes  
\be
s=\f{4\pi r^3_0}{2\kappa^2}\bigg[1+12 c_1-\f{q^2}{r^6_0}(18c_1+12c_2) \bigg]
\ee

giving the ratio 
\be
\eta/s=\f{1}{4\pi}\bigg[1-8~c_1+\f{q^2}{r^6_0}(4c_1+24c_2) \bigg],
\ee
which in the extremal limit gives
\be
(\eta/s)_{{\rm Extremal}}=\f{1}{4\pi}[1+48~c_2],
\ee
independent of $c_1$ as in the previous example.

 Using the holographic anomaly calculation, we can identify the coefficient $c_1$ with the central charges as 
\be
c_1=\f{1}{8}\bigg({\f{c}{a}-1}\bigg)
\ee

Using eq(\ref{a_c_result}), we get the constraint on $c_1$ as
\bea\label{bound_weyl_c_1}
&&{\rm For}~~~{\cal N}=0,~~~-\f{39}{248}~\leq ~3c_1~\leq~\f{3}{4},\nn
&&{\rm For}~~~{\cal N}=1,~~~-\f{1}{8}~\leq ~3c_1~\leq~\f{3}{8},\nn
&&{\rm For}~~~{\cal N}=2,~~~-\f{3}{40}~\leq ~3c_1~\leq~\f{3}{8}.
\eea

It would be interesting to find the restriction on $c_2$ using the anomaly calculation \cite{hs}.

\section{Solution to constraints from WGC and Central charges  }

Solving  eq(\ref{bound_c_1}) and eq(\ref{wgc}), we get solutions, which  has got both intervals and some isolated points. 
For ${\cal N}=0$, it is
\bea
&&c_1=\f{1}{12},~c_2=-\f{23}{576} ~{\rm and} ~ c_1=-\f{13}{144},~c_2=\f{731}{6912},\nn &&
-\f{23}{576}~\leq~c_2 ~\leq \f{13}{576},~ -\f{48 c_2}{23}~\leq~ c_1\leq \f{1}{12},\nn &&
\f{13}{576}~\leq~c_2 ~\leq \f{299}{6912},~ -\f{48 c_2}{23}~\leq~ c_1\leq \f{3-48 c_2}{23},\nn &&
\f{299}{6912}~\leq~c_2 ~\leq \f{731}{6912},~ -\f{13}{144}~\leq~ c_1\leq \f{3-48 c_2}{23},
\eea

For ${\cal N}=1$, it is
\bea
&&c_1=\f{1}{16},~c_2=-\f{23}{768} ~{\rm and} ~ c_1=-\f{1}{16},~c_2=\f{71}{768},\nn &&
-\f{23}{768}~\leq~c_2 ~\leq \f{23}{768},~ -\f{48 c_2}{23}~\leq~ c_1\leq \f{1}{16},\nn &&
\f{23}{768}~\leq~c_2 ~\leq \f{25}{768},~ -\f{1}{16}~\leq~ c_1\leq \f{1}{16},\nn &&
\f{25}{768}~\leq~c_2 ~\leq \f{71}{768},~ -\f{1}{16}~\leq~ c_1\leq \f{3-48 c_2}{23},
\eea

Finally for ${\cal N}=2$, it is
\bea
&&c_1=\f{1}{16},~c_2=-\f{23}{768} ~{\rm and} ~ c_1=-\f{1}{32},~c_2=\f{119}{1536},\nn &&
-\f{23}{768}~\leq~c_2 ~\leq \f{23}{1536},~ -\f{48 c_2}{23}~\leq~ c_1\leq \f{1}{16},\nn &&
\f{23}{1536}~\leq~c_2 ~\leq \f{25}{768},~ -\f{1}{32}~\leq~ c_1\leq \f{1}{16},\nn &&
\f{25}{768}~\leq~c_2 ~\leq \f{119}{1536},~ -\f{1}{32}~\leq~ c_1\leq \f{3-48 c_2}{23},
\eea

 For each case of ${\cal N}=0,~1,~2$, there exists solutions, where  $c_2$ is both positive and negative, but it is not a priori clear which set to choose and which set to ignore.

\end{document}